\documentclass[doublecol]{epl2}
\usepackage{graphicx}
\usepackage{amsmath}
\usepackage{amssymb}

\title{Elastic and inelastic collisions of interfacial solitons and
 integrability of two-layer fluid system subject to horizontal vibrations}
\shorttitle{Collisions of interfacial solitons and integrability
 in two-layer fluid system subject to horizontal vibrations}

\author{Denis S.\ Goldobin$^{1,2,3}$, Kseniya V.\ Kovalevskaya$^1$ \and Dmitry V.\ Lyubimov$^2$
 }
\shortauthor{D.\ S.\ Goldobin, K.\ V.\ Kovalevskaya, D.\ V.\ Lyubimov} 
\institute{$^1$~Institute of Continuous Media Mechanics, UB RAS,
         1 Academik Korolev street, Perm 614013, Russia \\
         $^2$~Department of Theoretical Physics,
         Perm State University, 15 Bukireva str., 614990, Perm, Russia \\
         $^3$~Department of Mathematics, University of Leicester,
         University Road, Leicester LE1 7RH, UK}


\abstract{
We study interfacial waves in a system of two horizontal layers of
immiscible inviscid fluids involved into horizontal vibrational
motion. We analyze the linear and nonlinear stability properties
of the solitons in the system and consider two-soliton collision
scenarios. We describe the events of explosive formation of sharp
peaks on the interface, which may presumably lead to the layer
rapture, and find that beyond the vicinity of this peaks the
system dynamics can be represented as a kinetics of a soliton gas.
 }

\begin{document}

\maketitle

\section{Introduction}
\begin{figure}[t]
\center{
 \includegraphics[width=0.43\textwidth]%
 {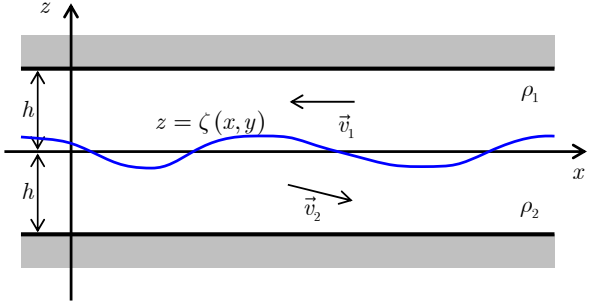}}

  \caption{
Sketch of a two-layer fluid system subject to harmonic
horizontal vibrations and the coordinate frame.
 }
  \label{fig_sketch}
\end{figure}

The experimental observations of the occurrence of steady wave
patterns on the interface between immiscible fluids subject to
horizontal vibrations were first reported by
Wolf~\cite{Wolf-1961,Wolf-1970}. Wolf also noticed the
opportunities for vibrational stabilization of the system states,
which are gravitationally unstable in the absence of vibrations,
and initiated exploration for these possibilities. The development
of a rigorous theoretical basis for these experimental findings
was contributed by the linear instability analysis of the flat
state of the
interface~\cite{Lyubimov-Cherepanov-1987,Khenner-Lyubimov-Shotz-1998,Khenner-etal-1999}
(in Fig.\,\ref{fig_sketch}, one can see the sketch of the system
considered in these works). It was found that in thin layers the
instability is a long-wavelength
one~\cite{Lyubimov-Cherepanov-1987}.
In~\cite{Khenner-Lyubimov-Shotz-1998,Khenner-etal-1999}, the
linear stability was analyzed for the case of arbitrary frequency
of vibrations.

In Wolf's experiments with horizontal vibrations~\cite{Wolf-1961},
the viscous boundary layer in the most viscous liquid was one
order of magnitude smaller than the layer thickness, meaning the
approximation of inviscid liquid to be appropriate. According
to~\cite{Lyubimov-Cherepanov-1987}, the layer is thin enough for
the marginal instability to be long-wavelength, when its
half-thickness $h<\sqrt{3\alpha/[(\rho_2-\rho_1)g]}$, where
$\alpha$ is the interface tension coefficient, $\rho_1$ and
$\rho_2$ are the light and heavy liquid densities, respectively,
and $g$ is the gravity. This critical layer thickness can be one
or two orders of magnitude larger than the thickness of the
viscous boundary layer, meaning the problem with long-wavelength
instability remains physically relevant for the case of inviscid
fluids. In the opposite limiting case, for a
viscosity-dominated system, the problem of pattern formation was
studied
in~\cite{Shklyaev-Alabuzhev-Khenner-2009,Benilov-Chugunova-2010}.
The case of dynamics of nearly-inviscid system is essentially
different from the purely dissipative dynamics reported
in~\cite{Shklyaev-Alabuzhev-Khenner-2009} for extremely thin
layers.

Until recently~\cite{Goldobin-etal-Nonlinearity-2014}, advances in
theoretical studies of the relief of interface or free surface
under high-frequency vibration fields were focused on the
quasi-steady profiles (e.g.,
\cite{Lyubimov-Cherepanov-1987,Zamaraev-Lyubimov-Cherepanov-1989}).
Within the approach
of~\cite{Lyubimov-Cherepanov-1987,Zamaraev-Lyubimov-Cherepanov-1989},
a kind of energy variational principle can be derived. This
principle was employed for calculation of the average profile
shape about which the interface trembles with small amplitude and
high frequency. This approach, however, does not allow considering
the pattern evolution and determining stability properties of the
quasi-stationary relief.
In~\cite{Goldobin-etal-Nonlinearity-2014}, the rigorous weakly
nonlinear analysis was applied for derivation of the governing
equations for large-scale (long-wavelength) patters below the instability threshold.
With these equations the family of solitons can be found in the
system. Remarkably, the standing solitons, which are the only
patterns that could be derived with the variational principle, are
always unstable. Thus, the governing equations we derived
in~\cite{Goldobin-etal-Nonlinearity-2014} provide for the first
time opportunity for a reliable and physically informative
theoretical analysis of nonlinear dynamics of the system.

With this letter, we will provide a comprehensive analysis of the
dynamics resulting from the nonlinear evolution equations for
long-wavelength patterns. For the soliton families we reported
in~\cite{Goldobin-etal-Nonlinearity-2014}, we will analyze
nonlinear stages of development of perturbations leading to either
an `explosion' or to falling-apart of an unstable soliton into
pair of stable solitons; the latter kind of behavior was
previously reported for soliton-bearing systems
in~\cite{Orlov-1983,Falkovich-Spector-Turitsyn-1983,Bogdanov-Zakharov-2002}.
The self-similar explosion solution agrees remarkably well with
the results of direct numerical simulation. Two integrals of
motion will be derived for the equations, corresponding to the
laws of conservation of mass and momentum in the virgin physical
system. With these integrals of motion, we will see that unstable
solitons can be represented by superpositions of pairs of stable
ones, while stable solitons are elementary, in a sense that they
can not be decomposed into any superpositions. The equivalence
between unstable solitons and certain pairs of stable solitons
suggests that stable soliton collisions can be `inelastic'. In
agreement with the latter, we will find that soliton collisions
can be either elastic or lead to an explosion; at the boundary
between elastic and explosive collisions, colliding stable
solitons coalesce into unstable ones. Finally, we will
see that the system dynamics can be completely represented as a
kinetics of a soliton gas and governed by the equation which is
known from~\cite{Bogdanov-Zakharov-2002} to be fully integrable
beyond the vicinities of the explosion sites.
Thus, we deal with the situation where the nonlinear dynamics of a
real physical system---which pertains to one of the classic
problems of fluid dynamics---can be fully integrated, and, even
more intriguingly, demonstrates features which are not very common
for soliton-bearing systems, such as decomposition of certain
solitons, possibility of inelastic collisions, etc.

\section{Governing equations for large-scale patterns}
With the standard multiscale method one can derive, that
large-scale ({\it or} long-wavelength) patterns in the
system of inviscid liquids are governed by the equation
system~\cite{Goldobin-etal-Nonlinearity-2014}
\begin{equation}
\left\{
 \begin{array}{rcl}
 \displaystyle
 B_0\frac{\rho_2+\rho_1}{\rho_\ast}T^2\frac{\partial\varphi}{\partial t}\!\!\!&
 \displaystyle =&\!\!\!
 \displaystyle \left[L^2-\frac{h^2}{3}\right]\frac{\partial^2\eta}{\partial x^2}
 \\[15pt]
  &&\displaystyle
  {}+\frac{3}{2h}\frac{\rho_2-\rho_1}{\rho_2+\rho_1}\eta^2
  +\frac{B_1}{B_0}\eta\,,\\[15pt]
  \displaystyle
 \frac{\partial\eta}{\partial t}\!\!\!&
 \displaystyle =&\!\!\!
 \displaystyle -h\frac{\partial^2\varphi}{\partial x^2}\,.
 \end{array}
\right.
\label{eq:dimensional}
\end{equation}
Here $\eta(x,t)$ is the non-pulsating part of the interface
displacement from the flat state, $\varphi(x,t)$ is the
non-pulsating part of the upper fluid flow
($\vec{v}_1=-\nabla\varphi+...$, $\vec{v}_2=\nabla\varphi+...$,
where ``$...$'' stands for the pulsing part of the flow and
smaller corrections); notice $\varphi(x,t)$ is independent of $z$,
since $\vec{v}_{1,2}$ are nearly constant along $z$. Reference
length $L=\sqrt{\alpha/[(\rho_2-\rho_1)g]}$, $\alpha$: surface
tension, $\rho_1$ and $\rho_2$: density of the upper and lower
fluids, respectively, $\rho_1<\rho_2$, $g$: gravity,  $h$:
unperturbed thickness of the layers, $T=L/b$: reference time,
$\rho_\ast$: reference fluid density ($\rho_\ast$ can be chosen as
convenient). Parameter $B$ is the dimensionless vibration
parameter;
\begin{equation}
 B\equiv\rho_\ast b^2/\sqrt{\alpha(\rho_2-\rho_1)g}=B_0+B_1\,,
\label{eq:B}
\end{equation}
where $b$ is the container vibration velocity amplitude; $B_0$ is
the linear instability threshold
\begin{equation}
 B_0=\frac{\rho_\ast(\rho_2+\rho_1)^3h}{2\rho_1\rho_2(\rho_2-\rho_1)^2}
 \sqrt{\frac{(\rho_2-\rho_1)g}{\alpha}}\,,
\label{eq:B0}
\end{equation}
and $B_1$ is the deviation from the stability threshold.

We consider the system dynamics slightly below the
linear instability threshold, i.e., for $B_1<0$ and
$|B_1|\ll B_0$. With rescaling
\begin{equation}
\begin{array}{c}
 x\to x\,L\sqrt{\frac{B_0}{(-B_1)}\left[1-\frac{h^2}{3L^2}\right]}\,,
 \quad
 \eta\to\eta\,h\frac{\rho_2+\rho_1}{\rho_2-\rho_1}\frac{(-B_1)}{B_0}\,,
 \\[10pt]
 t\to t\sqrt{\frac{\rho_2-\rho_1}{\rho_\ast}\frac{L^3B_0^3}{h\,b^2B_1^2}
 \left[1-\frac{h^2}{3L^2}\right]}\,,\mbox{ and}\\[10pt]
 \varphi\to\varphi\sqrt{\frac{\rho_\ast(\rho_2+\rho_1)^2}{(\rho_2-\rho_1)^3}
 \frac{L^3B_1^2}{h\,b^2B_0^3}\left[1-\frac{h^2}{3L^2}\right]}\,,
 \end{array}
\label{eq:rescaling}
\end{equation}
the governing equations~(\ref{eq:dimensional}) take
zero-parametric dimensionless form;
\begin{eqnarray}
\dot\varphi&=&\eta_{xx}+{\textstyle\frac{3}{2}}\eta^2-\eta\,,
\label{eq:dimensionless1}
\\
\dot\eta&=&-\varphi_{xx}\,.
\label{eq:dimensionless2}
\end{eqnarray}
Here subscripts denote the partial derivative with respect to the
specified coordinate.

The latter equation system can be recast as a `plus' Boussinesq
equation (BE);
\begin{equation}
\ddot{\eta}-\eta_{xx}+\left({\textstyle\frac{3}{2}}\eta^2+\eta_{xx}\right)_{xx}=0\,.
\label{eq:plBe}
\end{equation}
From the view point of dynamics, this equation essentially differs
from original Boussinesq equation B (BE~B) for waves in a shallow
water layer~\cite{Boussinesq-1872} or in a two-layer system
without vibrations~\cite{Choi-Camassa-1999}, which is
\begin{equation}
\ddot{\eta}-\eta_{xx}-\left({\textstyle\frac{3}{2}}\eta^2+\eta_{xx}\right)_{xx}=0\,.
\label{eq:BeB}
\end{equation}
The equation system
(\ref{eq:dimensionless1})--(\ref{eq:dimensionless2}) is rigorously
derived for the vicinity of the vibrational instability threshold
(e.g., one cannot consider the case of vanishing vibrations, when
departure from the threshold is finite, with this equation
system). On the contrast, the Boussinesq equation A
(BE~A)~\cite{Boussinesq-1872} with nonlinear term
$[(\varphi_x)^2+(1/2)\eta^2]$ in place of $(3/2)\eta^2$ in
Eq.\,(\ref{eq:BeB}) is derived for small dispersion and nonlinear
terms, i.e., only small-amplitude waves with velocity close to $1$
are quantitatively governed by BE~A. With assumptions required by
BE~A, one needs further to restrict consideration to the case of
the wave package moving in one direction for to set
$(\varphi_x)^2=(\dot\varphi)^2\approx\eta^2$ and obtain BE~B
(Eq.\,(\ref{eq:BeB})). Summarizing, the results on soliton waves
derived with BE~B are rigorous for waves in shallow water only for
the edge of the soliton spectrum and never rigorous for collisions
of contrpropagating solitons, while Eq.\,(\ref{eq:plBe}) is
rigorous for interfacial waves in the system subject to horizontal
vibrations.

The integrability of the `plus' BE was considered
in~\cite{Bogdanov-Zakharov-2002} where the $\partial$-dressing
method was employed for deriving multisoliton solutions. Bogdanov
and Zakharov~\cite{Bogdanov-Zakharov-2002} reported existence of
unstable solitons which can decay into pairs of stable solitons
and thoroughly treated bounded states of two and more `singular'
solitons of the form $\eta=-4/(x-x_0)^2$. For our system the
`singular' solitons cannot be considered as the long-wavelength
approximation is violated for them. As
Ref.~\cite{Bogdanov-Zakharov-2002} does not provide answers to
some significant questions and omit certain important scenarios of
the system dynamics, it will be more convenient to perform a
comprehensive analysis of the system dynamics, without employment
of a laborious $\partial$-dressing technique, and postpone
comparison of this analysis to~\cite{Bogdanov-Zakharov-2002} for
the Discussion section.

\section{Solitons}
Equation
system~(\ref{eq:dimensionless1})--(\ref{eq:dimensionless2}) admits
solutions in the form of propagating wave with time-independent
profile; $\eta(x,t)=\eta(x-ct)\equiv\eta(\xi)$,
$\varphi(x,t)=\varphi(x-ct)\equiv\varphi(\xi)$, where $c$ is the
wave propagation speed. For these waves $\partial_x=\partial_\xi$
and $\partial_t=-c\partial_\xi$, and
Eqs.\,(\ref{eq:dimensionless1})--(\ref{eq:dimensionless2}) yield
\begin{equation}
\textstyle
0=\eta_{\xi\xi}+\frac32\eta^2-(1-c^2)\eta\,.
\label{eq:s01}
\end{equation}
(Here we used the condition $\eta(\xi=\pm\infty)=0$.) The latter
equation admits the soliton solution
\begin{equation}
\eta_0(\xi)=\frac{1-c^2}{\displaystyle\cosh^2\big[\sqrt{1-c^2}\xi/2\big]}\,;
\label{eq:s02}
\end{equation}
for a given initial profile $\eta(x)$, the propagation direction
($+c$ or $-c$) is determined by the flow, $\varphi_\xi=\pm
c\eta(\xi)$ (cf.\ Eq.\,(\ref{eq:dimensionless2})). The family of
solitons is one-parametric, parameterized by the speed $c$ only.
Speed $c$ varies within the range $[0,1]$; the standing soliton
($c=0$) is the sharpest and the highest one, while for the fastest
solitons, $c\to1$, the width tends to infinity ($\propto
1/\sqrt{1-c^2}$) and the height tends to $0$ ($\propto [1-c^2]$).

Let us interpret these results in original dimensional
space--time. The spatial and temporal scales for dynamic
system~(\ref{eq:dimensionless1})--(\ref{eq:dimensionless2}) depend
on the deviation from the threshold $(-B_1)$ (see
rescaling~(\ref{eq:rescaling})). For a transparent interpretation
of the dynamics of patterns in original dimensional space--time,
one can consider solitons for the dimensional equation
system~(\ref{eq:dimensional}) and find equation of
form~(\ref{eq:s01}) with coefficient
\begin{equation}
G:=\frac{(-B_1)}{B_0}-c_{\mathrm{dim}}^2\frac{(\rho_2+\rho_1)\,\alpha^{1/2}}{h\,[(\rho_2-\rho_1)g]^{3/2}}
\label{eq:s03}
\end{equation}
ahead of the last term $(-\eta)$, where $c_{\mathrm{dim}}$ is the
dimensional speed of the soliton, with all the other coefficients
being physical parameters of the physical system under
consideration. Considering a given physical system with vibration
parameter $B$ as a control parameter, one can see that the shape
of a soliton is controlled by expression (\ref{eq:s03}). Hence,
the same interface inflection soliton can exist for different
deviation $B_1$, but the same $G$, which will be achieved by
tuning $c_\mathrm{dim}$; for larger negative deviation $(-B_1)$
from the instability threshold, the speed $c_\mathrm{dim}$ is
larger.

The family of solitons can be compared against the wave packages
of linear waves---small perturbation of the flat-interface state.
For small normal perturbations
 $(\eta,\varphi)\propto e^{i(kx-\Omega t)}$,
Eqs.\,(\ref{eq:dimensionless1})--(\ref{eq:dimensionless2}) yield
the dispersion relation $\Omega(k)=k\sqrt{1+k^2}$. The group
velocity is
\begin{equation}
v_\mathrm{gr}=\frac{d\Omega}{dk}=\frac{1+2k^2}{\sqrt{1+k^2}}\,,
\label{eq:s04}
\end{equation}
which varies from $v_\mathrm{gr}=1$ (for $k\to0$) to infinity (for
$k\to\infty$). Thus, any packages of linear waves travel with
higher velocity than the fastest solitons.

\begin{figure}[t]
\center{
 \includegraphics[width=0.475\textwidth]%
 {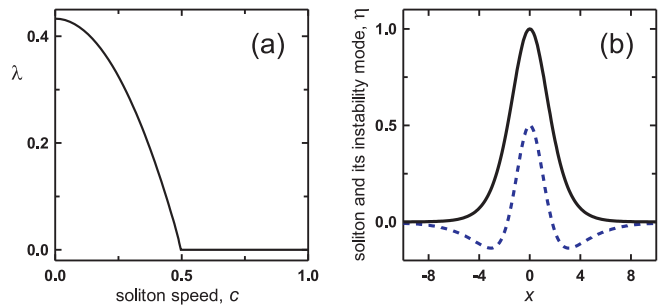}}

  \caption{
(a):~Exponential growth rate $\lambda$ of perturbations of the
soliton vs soliton speed $c$. (b):~Soliton (black solid line) and
its instability mode (blue dashed line) for $c=0$.
 }
  \label{fig_stability}
\end{figure}

\begin{figure*}[t]
\center{
 \includegraphics[width=0.85\textwidth]%
 {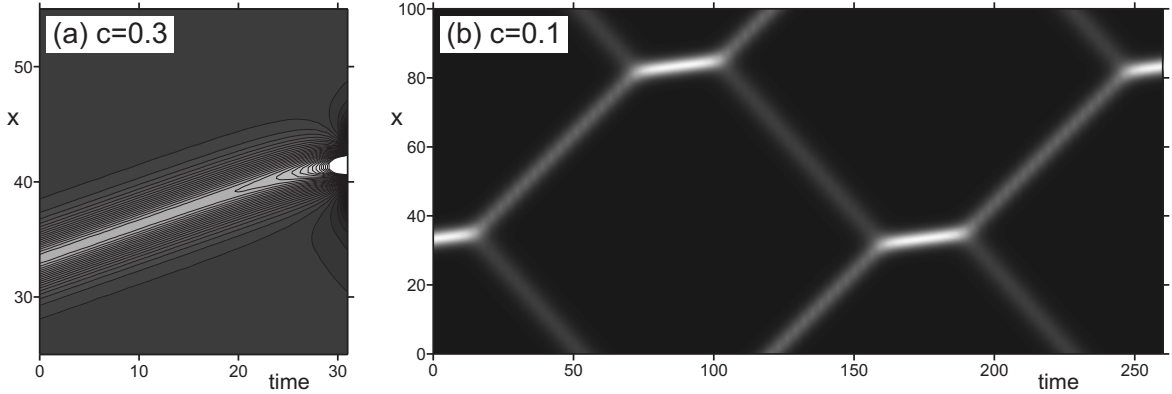}}

  \caption{
Two scenarios of nonlinear evolution of perturbations of unstable
solitons. (a):~Explosion of unstable soliton with $c=0.3$.
(b):~Falling-apart of unstable soliton with $c=0.1$ and
recoalescence of the stable decay products.
 }
  \label{fig_nonlin_instab}
\end{figure*}

\section{Stability of solitons (initial perturbations)}
The linear stability analysis for solitons in dynamic
system~(\ref{eq:dimensionless1})--(\ref{eq:dimensionless2})
reveals that the slow solitons, with $0\le c\le1/2$, are unstable
(see Fig.\,\ref{fig_stability}(a)), with one unstable degree of
freedom (in Fig.\,\ref{fig_stability}(b), one can see the
instability mode for $c=0$). The fast solitons, with $1/2<c\le1$,
are stable both linearly and to non-large finite perturbations.
Detailed analysis of the linear stability can be found
in~\cite{Goldobin-etal-Nonlinearity-2014}.

It is interesting to consider the nonlinear development of
perturbations of unstable solitons. Two possible scenarios
were encountered in direct numerical simulations:
\\
(i)~Explosive growth of perturbation and formation of an
infinitely sharp and high peak in a finite time;
\\
(ii)~Falling-apart of the unstable soliton into exactly two stable
ones.
\\
In Fig.\,\ref{fig_nonlin_instab}, one can see the development of
these scenarios.

Obviously, for the scenario~(i), after violation of the conditions
of the long-wavelength approximation, the dynamics will deviate
from the one dictated by
Eqs.\,(\ref{eq:dimensionless1})--(\ref{eq:dimensionless2}); still,
the formation of an sharpening of the interface with large
deviation from the flat state is certain. In the following we will
derive scaling laws for this explosion regime.

For the falling-apart of the unstable solution, we can observe in
Fig.\,\ref{fig_nonlin_instab}(b) that two fast solitons can then
collide and coalesce again into the same initial unstable soliton,
which will exist for awhile. The smaller perturbation of this
unstable soliton, the longer it exists before falling apart again.
It suggests that collisions of solitons can be `inelastic'. In the
text below, we will investigate the collisions of fast stable
solitons numerically and reveal analytical conditions for
coalescence of colliding solitons, elastic collisions and
explosions (notice, in Fig.\,\ref{fig_nonlin_instab}(b) we observe
collision not for an arbitrary pair of stable solitons but for the
products of decay of the unstable soliton).

Noteworthy, one can predict wether the initial infinitesimal
perturbation will lead to one or another scenario. It is
determined by the projection of the initial perturbation onto the
unstable mode. According to results of direct numerical
simulation, if one normalizes the instability mode
 $(e^{\lambda t}\eta_1(\xi),e^{\lambda t}\varphi_1(\xi))$
so that $\eta_1(0)>0$---cf.\ Fig.\,\ref{fig_stability}(b), where
the instability mode for the soliton with $c=0$ is plotted with
the blue dashed line---then the perturbation with a positive
contribution of $(\eta_1(\xi),\varphi_1(\xi))$ will lead to
explosion while the one with a negative contribution will lead to
the splitting.[\footnote{This can be intuitively expected from
Fig.\,\ref{fig_stability}(b) as well. Indeed, the addition of the
dashed profile to the soliton (solid line) with a positive weight
means shrinking of the interface embossment and increase of its
height, which is the beginning of explosion, while the subtraction
of the dashed profile corresponds to decrease of the middle peak
and further formation of two peaks on the sides of the main one,
which are `embryos' of two splitting products.}]

\begin{figure}[t]
\center{
 \includegraphics[width=0.475\textwidth]%
 {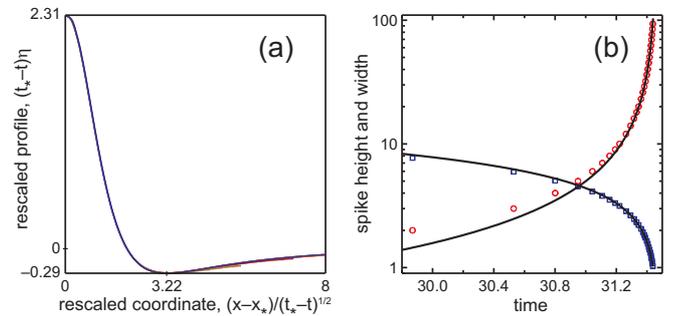}}

  \caption{
(a):~Profiles of the explosively growing spike from
Fig.\,\ref{fig_nonlin_instab}(a) for $(t_\ast-t)=0.1$, $0.2$,
$0.3$, $0.4$, $0.5$, $0.6$ (where $t_\ast$ is the time instant
when $\eta_\mathrm{max}\to\infty$) are nearly indistinguishable
after rescaling. (b):~The behavior of the spike width (blue
squares) and height (red circles) is compared to the scaling laws
$6.448\sqrt{t_\ast-t}$ and $2.309/(t_\ast-t)$, respectively,
dictated by Eq.\,(\ref{eq:e02}) with $n=-1$, $m=1/2$ for
$\eta\gg1$.
 }
  \label{fig_explosion}
\end{figure}

\section{Explosions}
For the explosion solution, field $\eta$ becomes large and the
term $(-\eta)$ can be neglected against the background of the term
$(3/2)\eta^2$ in Eq.\,(\ref{eq:dimensionless1}); therefore,
equation
system~(\ref{eq:dimensionless1})--(\ref{eq:dimensionless2}) can be
rewritten as
\begin{equation}
\ddot{\eta}\approx-\big(\eta_{xx}+{\textstyle\frac32}\eta^2\big)_{xx}\,.
\label{eq:e01}
\end{equation}
The last differential equation is homogeneous: it admits
self-similar solutions of the form
\begin{equation}
\eta(x,t)=t^nf(s),\qquad s=x/t^m\,,
\label{eq:e02}
\end{equation}
where $n$ and $m$ are to be determined from the condition that
Eq.\,(\ref{eq:e01}) yields a differential equation for $f(s)$
which is free from $t$ and $x$. After substitution~(\ref{eq:e02}),
Eq.\,(\ref{eq:e01}) reads
$$
\begin{array}{r}
n(n-1)f-m(2n-m-1)sf^\prime+m^2s^2f^{\prime\prime}
\qquad\\[10pt]
 =-t^{2-4m}\big[f^{\prime\prime}
 +\frac32 t^{n+2m}f^2\big]^{\prime\prime}\,,
\end{array}
$$
i.e.\ requires $m=1/2$ and $n=-1$. With these values of $n$ and
$m$ the equation for $f(s)$ reads
\begin{equation}
\textstyle
f^{\prime\prime\prime\prime}+\frac32(f^2)^{\prime\prime}
 +\frac14 s^2f^{\prime\prime}+\frac74 sf^\prime+2f=0\,.
\label{eq:e03}
\end{equation}
The last equation has unique solution neither diverging at $s=0$
nor nonvanishing at $s\to\pm\infty$. This solution is
\begin{equation}
f(s)=8\big(3\sqrt{2}-s^2\big)/\big(3\sqrt{2}+s^2\big)^2\,.
\label{eq:peak}
\end{equation} 
In Fig.\,\ref{fig_explosion}(a), one can see that the rescaled
profiles of the explosion solution from
Fig.\,\ref{fig_nonlin_instab}(a) become nearly indistinguishable
from the solution $f(s)$ to Eq.\,(\ref{eq:e03}) quite quickly.

While the solitons are running, the explosive solution is
standing. This can be seen as well in
Fig.\,\ref{fig_nonlin_instab}(a), where the pattern stops when the
explosion happens.

\begin{figure}[t]
\center{
 \includegraphics[width=0.490\textwidth]%
 {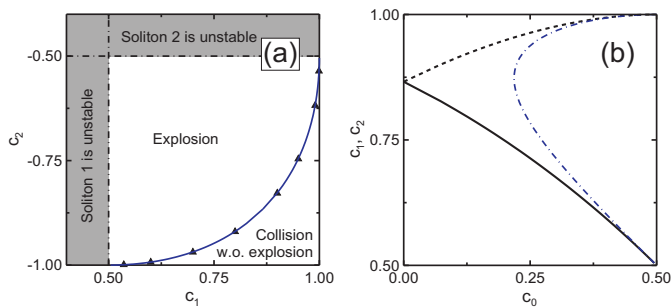}}

  \caption{
(a):~Collision map for contrpropagating solitons. On the blue
curve, colliding solitons coalesce; on the one side of the curve,
collisions result in explosion, while on the other side,
collisions are elastic. The results of direct numerical simulation
are plotted with triangles; the blue curve represents the solution
to algebraic system (\ref{eq:c01})--(\ref{eq:c02}) for soliton
coalescence. (b):~Speeds of coalescing solitons $c_1$ and $c_2$ on
the collision instability boundary are plotted versus $c_0$, the
unstable soliton speed forming as a result of coalescence. The
black curves correspond to collision of contrpropagating solitons.
The blue dash-dotted curve corresponds to the formal coalescence
solution for copropagating solitons, which is not observed in
direct numerical simulation; copropagating solitons exchange their
momentum from distance, without formation of a single peak
interface profile at any stage of the collision event.
 }
  \label{fig_collisions}
\end{figure}

\section{Integrals of motion}
As an integrable system, BE possesses infinite number of integrals
of motion. In particular, one can show the dynamic system
(\ref{eq:dimensionless1})--(\ref{eq:dimensionless2}) to possess
two following independent integrals of motion[\footnote{Indeed,
$\dot{I}_0=\int\dot{\eta}\,\mathrm{d}x=-\int\varphi_{xx}\mathrm{d}x=0$,
and
$\dot{I}_1=\int(\dot{\eta}\varphi_x+\eta\dot{\varphi}_x)\mathrm{d}x
=\int(\dot{\eta}\varphi_x-\eta_x\dot{\varphi})\mathrm{d}x
=\int(-\varphi_x^2/2-\eta_x^2/2-\eta^3/2+\eta^2/2)_x\mathrm{d}x=0$;
the integrals here vanish as integrals of $x$-derivatives of
functions vanishing at infinity.}]:
\begin{align}
I_0&\textstyle=\int_{-\infty}^{+\infty}\eta(x,t)\,\mathrm{d}x\,,
\label{eq:i01}
\\
I_1&\textstyle=\int_{-\infty}^{+\infty}\eta(x,t)\,\varphi_x(x,t)\,\mathrm{d}x\,.
\label{eq:i02}
\end{align}
The first integral is owned by the mass conservation law and the
second integral represents the momentum conservation law---thus,
the both conservation laws valid for the virgin fluid dynamical
system have their reflections in presented integrals of motion of
the system (\ref{eq:dimensionless1})--(\ref{eq:dimensionless2}).
These integrals will be useful for our following consideration.

For the soliton~(\ref{eq:s02}), one can find
\begin{align}
I_0[\eta_0(x-ct)]&\textstyle=\int_{-\infty}^{+\infty}\eta_0\,\mathrm{d}x
 =\sqrt{1-c^2}\,,
\label{eq:i03}
\\
I_1[\eta_0(x-ct)]&\textstyle=c\int_{-\infty}^{+\infty}\eta_0^2\,\mathrm{d}x
 =c\left[1-c^2\right]^{3/2}.
\label{eq:i04}
\end{align}
Here it is important, that soliton velocity $c$ in
Eq.\,(\ref{eq:i04}) is negative for solitons propagating to the
left.


\section{Two-soliton collisions}
While the system dynamics is integrable, the possibility of
explosions and the coalescence of decay products of an unstable
soliton (as seen in Fig.\,\ref{fig_nonlin_instab}(b)) suggest that
collision of stable solitons can be `inelastic'. The results of
direct numerical simulation of the collisions of pairs of solitons
are presented in Fig.\,\ref{fig_collisions}(a). Collisions of
copropagating solitons are always elastic and they exchange their
velocities during this collisions. Collisions of contrpropagating
solitons can be either elastic, when solitons are fast enough, or
lead to an explosion. With elastic collision, they exchange
velocities and effectively transpass through one another. At the
explosion boundary, solitons are found to coalesce, forming an
unstable soliton. It is interesting to consider this coalescence
and the decay of unstable solitons into two stable products from
the perspective of the motion integrals. Indeed, for solitons that
are at the distance from each other the profiles are nearly
mutually unaffected and
$I_j=I_j[\eta_0(x-c_1t,c_1)]+I_j[\eta_0(x-c_2t,c_2)]$, where
$c_2<0$, which should be the same as for the coalescence product
soliton of velocity $c_0$. These integrals are best to be recast
in terms of $Z_j\equiv\sqrt{1-c_j^2}$;
\begin{align}
Z_1+Z_2&=Z_0\,,
\label{eq:c01}
\\
Z_1^3\sqrt{1-Z_1^2}-Z_2^3\sqrt{1-Z_2^2}
 &=Z_0^3\sqrt{1-Z_0^2}\,.
\label{eq:c02}
\end{align}
For any $Z_0\in(1/\sqrt{2};1]$, which corresponds to
$c_0\in[0;1/2)$, i.e.\ an unstable soliton, a pair of solutions
$Z_1$ and $Z_2$ exist, both of which are in the semi-open interval
$(0;1/\sqrt{2}]$, i.e.\ correspond to stable solitons with
$|c_j|\in[1/2;1]$. For $Z_0<1/\sqrt{2}$ the equation
system~(\ref{eq:c01})--(\ref{eq:c02}) has no solution. Thus, one
obtains the soliton collision instability boundary
semi-analytically, with the solution to algebraic equation
system~(\ref{eq:c01})--(\ref{eq:c02}). This solution is plotted in
Fig.\,\ref{fig_collisions}(a) with the solid blue line and one can
see this result to match the results of direct numerical
simulation plotted with triangles. In
Fig.\,\ref{fig_collisions}(b) one can see relations between $c_0$
and the pair $c_1$ and $|c_2|<c_1$.

This result is quite interesting: half of solitons, with
$c\in[0,1/2)$, are unstable and can be represented as a
superposition of two solitons from the other half of them, with
$c\in[1/2,1)$. The solitons of the latter half are stable and
cannot be decomposed into other solitons. Moreover, the unstable
solitons are not merely the superposition of stable ones, they are
also the boundary of the basin of the system trajectories leading
to an explosive rapture of the upper layer.

\begin{figure}[t]
\center{
 \includegraphics[width=0.475\textwidth]%
 {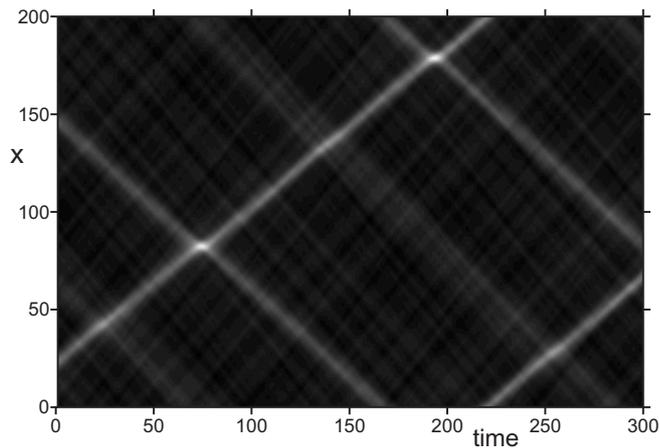}}

  \caption{
Sample evolution of the dynamic
system~(\ref{eq:dimensionless1})--(\ref{eq:dimensionless2}) with
arbitrary initial conditions: the dynamics turns out to be a
kinetics of a gas of stable (fast) solitons experiencing
collisions.
 }
  \label{fig_solitons_ensemble}
\end{figure}

\section{Discussion}
In Fig.\,\ref{fig_solitons_ensemble}, a sample of the system
dynamics from arbitrary initial conditions is presented in domain
$x\in[0;200]$ with periodic boundary conditions. One can see this
dynamics can be well treated as a kinetics of a gas of stable
solitons.

Let us compare the big picture of the system dynamics constructed
above with the results from~\cite{Bogdanov-Zakharov-2002} where
the dynamics of the `plus' Boussinesq equation was considered {\it
on manifolds of superpositions of finite number of solitons}. Two
subfamilies of stable and unstable solitons were revealed
in~\cite{Bogdanov-Zakharov-2002} as well, however their stability
was not considered with respect to arbitrary perturbations. While
the decay of an unstable soliton into pair of stable solitons was
reported with explicit analytical solutions, possibility of an
`explosion' of single unstable soliton was not reported. Thus, the
picture of scenarios of instability development was incomplete.
Although the formation of an explosion can be seen
in~\cite{Bogdanov-Zakharov-2002} for collisions of two unperturbed
unstable solitons, its universal asymptotic shape
(Eq.\,(\ref{eq:peak})) and scaling properties
(Eq.\,(\ref{eq:e02})) were not considered. For the problem of
two-soliton collisions, there were two general conclusions
in~\cite{Bogdanov-Zakharov-2002}: (1)~stable copropagating
solitons ``do not form singularities as a result of two-soliton
interaction'' (which is important for us, as we observe no
explosions and no coalescences for copropagating solitons) and
(2)~two-soliton interaction of unstable solitons necessarily leads
to formation of a singularity. In the light of the results of the
analysis of stability with respect to arbitrary perturbations, the
latter statement becomes less informative. The case of collision
of contrpropagating stable solitons, which yields us most
important results, was not addressed previously.
Ref.~\cite{Bogdanov-Zakharov-2002} stands as a prominent work in
the theory of solitons, presenting general multi-soliton solution
for a paradigmatic `plus' Boussinesq equation, the phenomenon of
decay of unstable solitons into pairs of stable ones, and forth
and back transformations of interacting solitons into bounded
states of singularities.

\section{Conclusion}
We presented a comprehensive analysis of the dynamics resulting
from the nonlinear evolution equations for long-wavelength
patterns in a system of two layers of immiscible inviscid fluids
subject to horizontal vibrations. In this system the standing and
slow solitons, $c<1/2$, are unstable, while the fast solitons,
$c\ge1/2$, are stable. For unstable solitons, nonlinear stages of
development of perturbations lead to either an `explosion' or to
falling-apart of an unstable soliton into pair of stable solitons.
The self-similar explosion soliton was derived and found to agree
well with the results of direct numerical simulation. Two
integrals of motion were obtained and employed for demonstrating
that unstable solitons can be represented by superpositions of
pairs of stable ones, while stable solitons are elementary. We
found that soliton collisions can be either elastic or lead to an
explosion; at the boundary between elastic and explosive
collisions, colliding stable solitons coalesce into unstable ones.
To conclude, beyond the vicinities of explosions, the system
dynamics is completely representable by a kinetics of a
soliton gas and the system is fully integrable. Coexistence of
explosion solutions and integrability for a real
physical system is quite remarkable.

\acknowledgements{
We are thankful to Dr.\ Maxim V.\ Pavlov, Dr.\ Takayuki Tsuchida,
two unknown reviewers, and Dr.\ Lyudmila S.\ Klimenko for their
useful comments on the paper and editor Prof.\ David Quere for his
work with the manuscript.} The work has been financially supported
by the Russian Science Foundation (grant no.~14-21-00090).

\bibliographystyle{eplbib}

\end{document}